# Investigation of biferroic properties in $La_{0.6}Sr_{0.4}MnO_3$ / 0.7 $Pb(Mg_{1/3}Nb_{2/3})O_3$ – 0.3 $PbTiO_3$ epitaxial bilayered heterostructures

Ayan Roy Chaudhuri[a] and S. B. Krupanidhi[b]
*Materials Research Centre, Indian Institute of Science, Bangalore 560 012, India*

P. Mandal and A. Sundaresan
*Chemistry and Physics of Materials Unit, Jawaharlal Nehru Centre for Advanced Scientific Research, Jakkur, Bangalore 560 064, India*

## Abstract:

Epitaxial bilayered thin films consisting of $La_{0.6}Sr_{0.4}MnO_3$ (LSMO) and 0.7 $Pb(Mg_{1/3}Nb_{2/3})O_3$ – 0.3 $PbTiO_3$ (PMN-PT) layers of relatively different thicknesses were fabricated on $LaNiO_3$ coated $LaAlO_3$ (100) single crystal substrates by pulsed laser ablation technique. Ferroelectric and ferromagnetic characteristics of these heterostructures confirmed their biferroic nature. The magnetization and ferroelectric polarization of the bilayered heterostructures were enhanced with increasing PMN-PT layer thickness owing to the effect of lattice strain. Dielectric properties of these heterostructures studied over a wide range of temperature under different magnetic field strength suggested a possible role of elastic strain mediated magnetoelectric coupling behind the observed magneto-dielectric effect in addition to the influence of rearrangement of the interfacial charge carriers under an applied magnetic field.

[a] Electronic mail: ayan@mrc.iisc.ernet.in
[b] Author to whom correspondence should be addressed; FAX: +9180 2360 7316
   Electronic mail: sbk@mrc.iisc.ernet.in

## Introduction:

Multiferroic materials (MFs) with co-existing ferroelectricity and magnetism have enjoyed a flurry of studies in recent years due to fundamental scientific interest and significant technological promises for their potential applications in future generation microelectronic devices, such as sensors, transducers, memory devices etc. The coexistence of ferroelectric (FE) and ferromagnetic (FM) properties in MFs and a coupling between them order parameters have been predicted to be useful in designing novel devices with parametric values and flexibility[1]. But the scarcity of single phase intrinsic MFs materials combined with their very feeble magnetoelectric (ME) response at room temperature have resulted in the realization of such multifunctional devices hitherto unachieved. The zest for understanding the mechanism of MF coupling and achieving substantial ME response have intrigued researchers worldwide towards alternative approaches to synthesize artificial magnetoelectric multiferroic materials. Various approaches have been made to design and synthesize artificial multiferroic structures. One of them is doping either a magnetic impurity in a ferroelectric host or a ferroelectric impurity in a magnetic host, or designing composites with ferroelectric and ferromagnetic hosts. Various bulk ME composites have been developed consisting of a FE constituent [e.g. $BaTiO_3$ (BT), $Pb(Zr_xTi_{1-x})O_3$ (PZT) etc.] and a FM constituent [e.g. $CoFe_2O_4$ (CFO), $NiFe_2O_4$ (NFO), Terfenol-D etc.]. In the BT/CFO bulk composite, the magnetic field induced dielectric response surpassed the values obtained from any single phase MF material by one order of magnitude[2,3]. The ME behaviour of such bulk composites generally depends on their microstructure and the coupling across the interface of the FE and the FM constituents[4,5]. More recently attention has been given to bilayers and multilayers of such composites where the induced ME effect arises as the product property of a magnetostrictive and a piezoelectric compound. These layered composites are especially promising due to their low leakage current and superior poling properties[6-8]. However such layered composite materials also suffer from several limitations such as poor mechanical coupling between layers due to non epitaxial nature of the interfaces, impurities arising from interfacial ion diffusion, lack of scaling possibilities etc. To overcome these difficulties involved with the bulk or layered composites, significant efforts have been devoted in designing ME nanostructures since they, especially ME thin films can easily undergo on-chip integration in microelectronic devices. The bilayers, superlattices and nano-composite thin film heterostructures combining FE and FM phases might have stronger feasibilities to overcome the difficulties associated with the bulk materials. These heterostructures have also exhibited stronger room temperature ME coupling compared to the single phase MFs[9,10]. While the superlattice approach has been investigated in great detail with different material-substrate combination to realize elastic strain mediated ME response[11-16], the bilayered heterostructures (BLs) have also found their niche in the ongoing MF research activities[17-22]. Recently electrical control of magnetism has been demonstrated in a bilayered thin film of $BiFeO_3$ and CoFe, where the interfacial exchange coupling between the $BiFeO_3$ and CoFe layers switches the magnetization of the FM CoFe layer on application of an electric field[23]. Furthermore, a prototype ME read head based on magnetic control of dielectric polarization has also been demonstrated by Zhang *et al.* utilizing BT/CFO and BT/NFO epitaxial bilayered thin films[24]. Our previous work[20] demonstrated that the dielectric properties of $La_{0.6}Sr_{0.4}MnO_3$ (LSMO)/$0.7Pb(Mg_{1/3}Nb_{2/3})O_3$-$0.3PbTiO_3$ (PMN-PT) epitaxial BLs can be influenced by an applied magnetic field. PMN-PT thin films having a very high piezoelectric coefficient[25] can be expected to exhibit substantially higher ME response when combined with magnetostrictive LSMO in thin film heterostructures. The strain states and therefore the functional properties of such BLs depend strongly on their microstructures as well as on the relative thickness of the constituent layers. In the present study we have chosen LSMO/PMN-PT BLs having different thickness ratios of the constituent layers. Structural, ferroelectric, ferromagnetic and magneto-dielectric properties of these heterostructures have been studied and discussed in relation to the lattice strain and interfacial charge dominated phenomena.

## Experimental:

Phase pure ceramic targets of the materials were prepared using precursors supplied by Aldrich Chemicals (99.9%). Thin film capacitors were fabricated by pulsed laser deposition (PLD), using a 248 nm KrF excimer laser (Lambda Physik COMPex) on single crystal LaAlO$_3$ (LAO) substrates placed at a distance of 4 cm from the targets to get a fluence of ~3.1 J/cm$^2$. The base pressure of the chamber was brought down to <1× 10$^{-6}$ m.bar prior to each deposition. A LaNiO$_3$ (LNO) bottom electrode layer of ~ 50 nm thickness was deposited before depositing the functional ferroic layers under 0.39 m bar Oxygen pressure, a substrate temperature of 700 °C and pulse repetition rate of 3 Hz. PMN-PT / LSMO heterostructures were thus deposited at a substrate temperature of 700 °C under oxygen partial pressure of 0.13 m. bar and pulsed repetition rate of 3 Hz for the LSMO layer and 5 Hz for the PMN-PT layer. The fabricated heterostructures were brought back to room temperature immediately after deposition with a cooling rate of 13 °C min$^{-1}$. The resulting BLs can be represented as PMNPT/LSMO/LNO/LAO. Four different BLs have been considered consisting of different thickness ratio of PMN-PT and LSMO. The LSMO layer thickness was varied between 20 nm-80 nm whereas the PMN-PT layer thickness was varied between 80 nm-140 nm keeping the total thickness of the BLs constant at ~ 160 nm. The thicknesses of the BLs were measured by cross sectional scanning electron microscopy (Quanta). In the present article the bilayered samples are represented as x/y, where x is the thickness of the PMN-PT layer (in nm) and y is the thickness of the LSMO layer (in nm). The four heterostructures thus can be represented as 80/80, 100/60, 130/30 and 140/20. Circular gold dots of diameter ~ 600 µ deposited by thermal evaporation using a shadow mask were used as top electrodes.

For crystallographic and epitaxial characterizations of the heterostructures, $\theta$-$2\theta$ and Phi ($\Phi$) scans were performed using a Bruker D8 Discover X Ray diffractometer (Cu $K_\alpha$, $\lambda = 0.15418$ nm). The surface morphologies of the BLs were investigated by contact mode Atomic Force Microscopy (AFM) (Veeco, CP II) studies. The dc magnetization measurements were performed using a vibrating sample magnetometer (VSM) by placing the sample in a physical properties measurement system (PPMS) (Quantum Design, USA) with the magnetic field parallel to the film plane. A Radiant Technology Precision ferroelectric workstation was used to measure the room temperature ferroelectric polarization hysteresis (*P-E*) at different frequencies and applied voltages. In order to measure the dielectric response under an applied magnetic field, the samples were mounted on a sample holder inserted in close cycle cryo-cooled magnet and connected to an Agilent 4294A impedance analyzer using co-axial compensated cables. For all the electrical measurements current perpendicular to the plane geometry has been used.

## Results and Discussion:

**a) Structural characterization**

Figure 1 shows the X-Ray diffraction pattern of the four BLs in the θ-2θ geometry. Appearance of only the (00*l*) peaks of PMN-PT and LSMO confirms highly oriented growth of the thin films. The epitaxial "cube on cube" growth of the heterostructures has been established by the Phi scan measurements performed on the substrate and the thin films which has been published elsewhere[20]. The *c*-axis lattice parameter calculated for PMN-PT from the (002) peak varied between 4.0829 Å to 4.0879 Å in these BLs indicating that the PMN-PT ($a$ = 4.025 Å) layer is under compressive in plane stress. On the other hand the *c*-axis lattice parameter of the LSMO ($a_{pc}$ = 3.87 Å) layer varied from 3.8827 Å for 80/80 heterostructure to 3.8711 Å for the 130/30 heterostructure. The decrease in the *c*-axis lattice parameter of LSMO with decreasing LSMO layer thickness accompanied by the increasing PMN-PT layer thickness indicated possible effect of the tensile in plane stress on the LSMO layer exerted by the PMN-PT layer. On increasing the PMN-PT layer thickness the tensile stress on the LSMO layer might increase thereby reducing its *c*-axis lattice parameter in order to keep the volume of the unit cell unchanged.

Figure 2 (a-d) shows the surface topography of the BLs. All the heterostructures exhibited dense surface morphology consisting of highly oriented grains. The morphology analysis revealed that the rms surface roughness of the BLs is a function of the film layer thickness. Figure 3 shows the rms roughness and the average in plane grain size of the heterostructures. The 80/80 film was found to have rms roughness of ~ 3.2 nm which was the highest among all the heterostructures considered. AFM studies were also performed on single layer LSMO thin films of different thicknesses and it was found that the surface roughness of the LSMO thin films increased from ~ 0.8 nm to ~ 3 nm with increasing thickness between 20 nm – 80 nm. Therefore the high value of roughness of the 80/80 bilayered thin film could be attributed to the rough interface between the PMN-PT and LSMO layer owing to the rough LSMO surface. For other heterostructures the rms roughness increased with increasing the PMN-PT layer thickness. The 80/80 heterostructure consists of grains with average in plane grain size of ~60 nm. With increasing PMN-PT layer thickness a layer of larger grains with an average grain size of ~100-120 nm appears on top of the smaller grain layer. However the average in plane grain size of the smaller grain layer beneath the larger grains remained almost independent of the PMN-PT layer thickness in all the heterostructures. Such microstructural features similar to those previously reported for epitaxial thin films of different perovskite oxides[26,27] could be indicative of the Stranski – Krastanov (SK) type growth mechanism of these heterostructures.

**B) Ferroelectric characterization**

Figure 4(a) shows room temperature (RT) *P-E* loops of the 140/20 heterostructure at different probing frequencies ranging between 200 Hz to 2 kHz. Very weak frequency dependence and saturated nature of the *P-E* loops suggested that the polarization response is intrinsic to the material under study and does not arise from any extrinsic effect, such as leakage current in this measured range of frequency. Figure 4(b) shows the variation of the remnant polarization ($P_R$) and the coercive voltage ($V_C$) as functions of applied voltage from 5 to 28 V with 1 kHz frequency. The $P_R$ and $V_C$ approached saturation with increasing voltage beyond 18 V which also supports the intrinsic FE response of the BLs. Such robust FE response was observed in the entire range of temperature between 20K and 300K[20]. All the BLs under study exhibited similar *P-E* response under an applied electric field at different probing frequencies. RT polarization response of the four BLs under a maximum applied electric field of 1250 kV/cm at 1 kHz has been shown in figure 5 (a). All the samples exhibited well saturated hysteresis. The spontaneous polarization ($P_S$) and $P_R$ values of the heterostructures increased from 26.3 μC/cm$^2$ and 13.4 μC/cm$^2$ respectively for the 80/80 heterostructure to 43.2 μC/cm$^2$ and 15.7 μC/cm$^2$ for the 140/20 heterostructure whereas the coercive field ($E_C$) followed a trend opposite to that of the polarization by decreasing from 257.6 kV/cm to 140 kV/cm. Figure 5 (b) summarizes the polarization and coercive field values in the BLs as a function of PMN-PT layer thickness ($d_{PMN-PT}$). The observed change in the polarization and coercive fields in these heterostructures can have more than one reason. Firstly the increase in the out of plane lattice parameter of PMN-PT from ~ 4.08 Å for 80/80 heterostructure to ~ 4.09 Å for 140/20 heterostructure indicated increased in-plane compressive stress on the PMN-PT layer with increasing thickness in these BLs. Compressive in plane stress is known to increase the polarization in epitaxial FE thin films[28,29]. In addition to this there can be the effects of space charge[30], depolarization[31], oxygen vacancies[32] and dielectric passive layers[33,34] at both Au/PMN-PT and PMN-PT/LSMO interfaces which determine the overall response of the heterostructures under an applied electric field. Moreover the possible defect sites present at the interface between lattice and polarization mismatched materials PMN-PT and LSMO can act as FE domain pinning centres[35,36] which in addition to the other effects mentioned can reduce the FE polarization and increase the $E_C$ in the thin film heterostructures. The interface dominated effects are more pronounced in case of films with reduced FE layer thickness. But, for a heterostructure system involving an intrinsically disordered FE material like PMN-PT and an interface of very complex nature with a manganite material it is difficult to assign the observed response to any particular mechanism unambiguously. Therefore, the observed FE response from the PMN-PT/LSMO BLs could be a collective effect of all the factors explained above.

**C) Ferromagnetic characterization**

Ferromagnetic behaviour of all the four different BLs have been studied within the temperature range 10 K-300 K with a sweeping dc magnetic bias of -3 kOe- +3 kOe applied parallel to the film plane. Figure 6 (a-c) shows the M-H response of the four BLs at three different temperatures. All the heterostructures possess a FM characteristic over the entire range of temperature which is evident from the nature of their M-H response. It was observed that maximum magnetization ($M_S$) value decreases with increasing thickness of LSMO layer from 140/20 to 80/80, whereas the coercive field ($H_C$) exhibited an increasing trend. The $M_S$ and $H_C$ values of the heterostructure as a function of LSMO layer thickness have been plotted in figure 6 (d). The trend in $M_S$ and $H_C$ values in the BLs as a function of LSMO layer thickness followed a trend opposite to that of the single layer LSMO thin films. Many authors have attributed the increase in $M_S$ and decrease in $H_C$ as a function of increasing film thickness in case of mangantie thin films collectively to the release of elastic strain with increasing film thickness[37], structural distortion, compositional inhomogeneity near the film-substrate interface[38] and magnetic dead layer effect[39,40]. In order to explain the behaviour observed in case of the PMN-PT/LSMO bilayers, let us consider first the architecture of these heterostructures. A schematic diagram of the PMN-PT/LSMO bilayered heterostructures under epitaxial strain has been shown in figure 7. The pseudocubic lattice parameter of LSMO ($a_{pc}$ = 3.87Å) is larger than the LAO substrate ($a$ = 3.79Å) and smaller than PMN-PT ($a$ = 4.025Å). Therefore in the BLs the LSMO layer experiences an in plane compressive stress induced by the substrate, and a simultaneous tensile stress induced by the PMN-PT layer. The increasing thickness of the PMN-PT layer from 80/80 structure to the 140/20 structure coupled with the decrease in the thickness of LSMO layer resulted in the increase of effective in plane tensile stress on the LSMO layer which was evident from the change in the out of plane lattice parameters of LSMO. The in-plane lattice parameter of LSMO thin film can change substantially under the simultaneous effect of compressive and tensile strain. Lee *et al.* in their report on microstructural and magneto-transport properties of LSMO thin films under simultaneous compressive and tensile strain demonstrated that the *c*/*a* ratio (~ 1.008) of LSMO thin films grown on LAO substrate decreases (~ 0.996) on introducing a $BaTiO_3$ ($a$ = 4.033Å) layer in the heterostructure, which exerts an in plane tensile stress on the LSMO layer[41]. In the present study the out of plane lattice parameter of LSMO for the 130/30 heterostructure ($c_{LSMO}$ = 3.871 Å) was found to be smaller than the out of plane lattice parameter (3.94 Å) of a 32 nm thick LSMO single layer film grown on LAO. This observation strongly indicates the effect of in plane tensile stress on the LSMO layer exerted by the PMN-PT layer in these BLs. The improvement of magnetization in the BLs with increasing volume fraction of the PMN-PT layer can therefore be attributed to the increased in plane lattice strain on the LSMO layer. LSMO thin films under tensile stress are known to exhibit an in plane easy axis of magnetization[41,42]. The tensile strain can effectively change the Mn-O-Mn bond angle in the distorted vertex sharing $MnO_6$ octahedra of the LSMO layer thereby reinforcing the double exchange FM interaction in the in plane direction of the BLs[41]. The drop in $M_S$ and increase in the $H_C$ values of these heterostructures on increasing the LSMO layer thickness could be attributed to presence of defect states at the PMN-PT/LSMO interface which can give rise to various effects such as strong domain wall pinning, spin disorder, orbital ordering, etc. The occurrence of defect states can have different origins. Firstly study of the surface topography of single layer LSMO thin films conducted by AFM revealed that the surface roughness of the LSMO thin films increased with film thickness (rms/nm = 0.8; 1.5, 2.2, 3, for $d_{LSMO}$/nm = 20; 30; 60; 80), which was in agreement with the observation reported by Steenbeck et al[43]. Such an increase in surface roughness can result in a rough interface between the PMN-PT and LSMO layers in the BLs which in effect can give rise to various interfacial defect states. Rough interfaces coupled with the decreasing PMN-PT layer thickness can reduce the effective tensile stress on the LSMO layers which in turn can result in reduction of the in plane magnetization values. Added to this the possible FM domain wall pinning by the defect sites could be attributed to the increase in $H_C$. Moreover, in the case of $La_{1-x}Sr_xMnO_3$, A-type antiferromagnetic state appears for x>0.5 in bulk samples[44], which in thin films can expand to a lower doping region under epitaxial strain[45]. In this context, LSMO with x=0.4 considered in the present study, is positioned near the boundary between the double-exchange FM and super exchange antiferromagnetic phases. Therefore if any charge transfer occurs between LSMO and PMN-PT through the interface, the marginal FM ordering in the LSMO layer may be destabilized. However the exact mechanism(s) behind these observations remains elusive without detailed information about the spin state, orbital occupancy of the electrons and microstructural characteristics of the PMN-PT/LSMO interfaces. Therefore, within the scope of the present study the observed FM behaviour from the BLs could be attributed to the collective or competitive effect of all the factors discussed above.

**D) Magneto-dielectric characterization**

Our observations revealed the influence of PMN-PT layer on the magnetic properties of these biferroic BLs and indicated a possible interfacial strain coupling between the piezoelectric PMN-PT and magnetostrictive LSMO layers. In order to investigate the manifestation of any elastic strain mediated ME coupling in these BLs their dielectric properties have been studied under different magnetic fields over a wide range of temperature between 20 K and 300 K. Since dielectric constants of FE materials are functions of temperature, at every temperature and magnetic field the system was stabilized before performing the measurements in order to avoid any experimental artifact. The change in dielectric response from the samples under magnetic field has been expressed in terms of magnetocapacitance (MC) in this article. MC, defined as MC(%) = $100 \times [C(H,T)–C(0,T)]/C(0,T)$, where $C(H,T)$ represents the capacitance at a magnetic field H and a temperature T, was calculated from the capacitance measured under different magnetic fields. Before investigating the MC of the heterostructures, dielectric response of single layer epitaxial PMN-PT thin films were studied under identical conditions. Figure 8 (a-d) shows the dielectric response of a single layer PMN-PT thin film under zero and 1T magnetic field measured at four different temperatures. No detectable change in the capacitance was observed over the entire range of temperature (20 K-300 K) under the applied magnetic field of 1 T. In order to further investigate the magnetic field dependence of dielectric response of the PMN-PT single layer thin film, the magnetic field strength was varied between 0 T and 3 T at all the four temperatures, but no significant change was observed. Figure 9 shows the MC response of the single layer PMN-PT thin films at different temperatures as a function of magnetic field strength The maximum MC value observed in case of single layer PMN-PT thin films was ~ $4.5 \times 10^{-4}$ % over the entire range of temperature which can be considered negligible for any practical purpose. MC values observed with this order of magnitude were therefore considered as experimental artifact and not the sample response. Figure 10(a-d) shows the representative capacitance vs. frequency response of a BL (140/20) under 0 T and 1 T magnetic fields at different temperatures. On applying the magnetic field a distinct change in capacitance in the low frequency range (<100 kHz) was observed at all the temperatures. Dielectric properties of all the BLs considered in the present study exhibited identical dependence on applied magnetic field. The MC calculated from the capacitance response of the bilayered heterostructures measured under a magnetic field of 1 T is shown in figure 11. MC of the BLs increased with increasing the temperature from 20 K, attained a maximum value of ~ 1-1.5 % in the temperature range of 175 K-190 K in different heterostructures, followed by a decrease to < 0.2% at 300 K. The trend of magneto-dielectric response was found to be qualitatively similar to that reported for MF $BiFeO_3$ thin films[46]. All the BLs exhibited an increase in MC with increasing magnetic field strength. A representative MC vs. magnetic field response obtained from 140/20 BL at 20 K has been shown in figure 12.

To explain the observed MC characteristics of the heterostructures elastic strain effect at the interfaces between PMN-PT and LSMO layers can be considered. In such cases the contribution of elastic energy to the total free energy of an epitaxial heterostructure under mechanical equilibrium can be expressed as[47]

$$\int_{-h_e/2}^{h_e/2} (\varepsilon^{0E} - \varepsilon^{Et} + \varepsilon^S)^2 dz + \frac{1}{2} C^M \int_{-h_m/2}^{h_m/2} (\varepsilon^{0M} - \varepsilon^{Mt} + \varepsilon^S)^2 dz + \frac{1}{2} HC^0 (\varepsilon^S)^2 \qquad (1)$$

Where, $\varepsilon^{0E}$ and $\varepsilon^{0M}$ are the misfit strains relative to the substrate in the FE and FM layers respectively. $\varepsilon^S$ is the induced elastic strain in the substrate layer and $C^E$, $C^M$ and $C^O$ are the corresponding elastic moduli. $\varepsilon^{Et}$ and $\varepsilon^{Mt}$ are the eigenstrains of the FE and FM transformations respectively, from which the piezoelectric and the magnetostrictive forces originate. $h_e$, $h_m$, and $H$ correspond to the respective thicknesses of the FE, FM layers and the substrate. Any perturbation to either of the eigenstrains on application of an external magnetic or electric field requires the other eigenstrain to change its value in order to maintain the mechanical equilibrium and system integrity. Observation of MC under an applied magnetic field can thus have this origin. At low temperature, under an applied magnetic field the mechanical deformation in the magnetic layer is due to Joule magnetostriction, which is caused by domain wall motion and domain rotation. The temperature dependence of MC suggests that at low temperature magnetic domain rotation might have least influence on the dielectric response of the heterostructure, which could be due to the substrate clamping effect. Thus the observed MC effect could be attributed primarily to the volume magnetostriction of the LSMO layer. Very small magnitude of the volume magnetostriction at low temperature results in the observed lower value of MC. Srinivasan *et al.* showed that the magnetostriction of LSMO decreases with increasing temperature beyond 200 K[6]. The decrease in MC at higher temperature could be associated with the weakening of the strain coupling owing to the decrease of magnetostriction of the LSMO layers. These observations suggest that strain mediated elastic coupling at the PMNPT/LSMO interface could result in the observed MC response. However the magnetostriction and piezoelectric properties of the heterostructures could not be characterized in the present study without what the observations remain qualitative. The dependence of MC on the direction of applied magnetic field is currently being investigated to elucidate any possible relation between magnetic field direction and the strength of strain mediated ME coupling.

Although the MC effect observed in the epitaxial BLs is suggestive of strain mediated ME coupling, the very small magnitude of MC is indicative of strong substrate clamping effect which makes the in plane mechanical deformation of the thin film layers difficult under an external applied magnetic/electric field. In such cases the MC might be related to the structural imperfections. In case of epitaxial thin films which are not perfectly flat, small epitaxial mesas might exist which can strain more compared to the large continuous film under any applied external stress. Possible existence of such epitaxial mesas can also be responsible for the observed MC behaviour in the present study. The MC observed in such artificial heterostructures can also have origin different from strain coupling. Localization of free charges at interfaces has been found to be responsible for the MC response observed in different materials, such as $CaCu_3Ti_4O_{12}$, p-n junction diodes etc. which do not otherwise possess any FM constituent and hence are not MF in nature[48]. However the localized charges may be of different origin and the interfaces of different shapes and nature. In the present study the BLs involve interfaces between two lattice and polarization mismatched materials PMN-PT and LSMO. Such interfaces can be the source of several charged defect states owing to the strain fields and uncompensated polarization. The interfacial defect states can include shallow trap sites, oxygen vacancies etc. The work functions of dielectric PMN-PT and ferromagnetic LSMO being different, band bending may occur at their interfaces giving rise to interfacial charge depletion (space charges). In such cases the heterostructures can exhibit an interface dominated dielectric response and can be described by the Maxwell -Wagner (MW) capacitor model. On application of a magnetic field the magnetic domains in the LSMO layers get oriented along the filed directions and consequently alter the charges at the interfaces eventually leading to an alteration of the dielectric response at the interface which can result in the observed MC behaviour. Moreover the resistance of the magnetoresistive LSMO layer and hence that of the interfacial layer between LSMO and PMN-PT might change on applying the applied magnetic field, which can influence the measured dielectric permittivity of the samples. Therefore along with the lattice strain driven ME coupling magnetoresistance (MR) combined with the MW effect can thus provide a parallel mechanism for MC effect[49]. Under the framework of interface dominated dielectric response and the MW type relaxation mechanism, the realization of low frequency MC response can be accounted for as follows. At high frequencies the mobile charge carriers (e.g. oxygen vacancy) with higher relaxation times cannot respond to the applied field, so that the measured capacitance represents simply that of two insulating capacitors in series. At low frequencies, on the other hand the charge carriers in the low resistivity layer do respond so that most of the field drops across the layer with higher resistivity and thus the apparent decrease in dielectric thickness result in an increased capacitance. Since the charge carriers can participate in the dielectric response at the low frequencies, the effect of magnetic field on them can be held responsible for the observed MC effect. The ac and dc conduction studies performed on these BLs as a function of temperature revealed the existence of charge depleted interfaces and trap sites of different activation energies which also support possible interfacial charge carrier dominated MC response in these BLs.

## Conclusions:

To summarize, biferroic heterostructures consisting of LSMO/PMN-PT bilayers with different volume fraction of the FM and FE phases have been fabricated epitaxially on LNO coated LAO (100) substrates by pulsed laser ablation technique. The lattice strain, surface roughness and the grain sizes was varied as a function of individual layer thickness without incurring any impurity phase or incoherency in the heterostructures. All the BLs exhibited desired ferroelectric and ferromagnetic properties. Enhanced magnetization owing to elastic strain was observed in these heterostructures compared to the magnetization of LSMO single layer thin films of equal thickness. The enhancement in magnetization was more significant as the PMN-PT layer thickness increased indicating the influence of increased lattice strain on LSMO by the PMN-PT layer. The dielectric characteristics under an applied magnetic field observed as a function of frequency and temperature indicated possible strain mediated ME coupling along with the interfacial charge dominated magneto-dielectric response in these artificial biferroic heterostructures.

Figure captions:

Figure 1. X-Ray diffraction patterns of LSMO/PMN-PT biferroic heterostructures.

Figure 2. 2D AFM micrographs of (a) 80/80, (b) 100/60, (c) 130/30 and (d) 140/20 LSMO/PMN-PT bilayered heterostructures.

Figure 3. RMS surface roughness and average in-plane grain sizes of the LSMO/PMN-PT bilayered heterostructures. The solid lines are guides to the eyes.

Figure 4(a). Room temperature *P-E* hysteresis loops of 140/20 LSMO/PMN-PT heterostructure at different probing frequencies.

Figure 4(b). ±$P_R$ and ±$V_C$ values of the 140/20 heterostructure at different applied voltages.

Figure 5 (a). Room temperature *P-E* hysteresis loops of different LSMO/PMN-PT heterostructures.

Figure 5(b). $P_R$, $P_S$ and $E_C$ as a function of PMN-PT layer thickness in the bilayered heterostructures.

Figure 6. M-H response of LSMO/PMN-PT bilayered heterostructures at (a) 10 K, (b) 100 K (c) 300 K and (d) $M_S$ and $H_C$ of the bilayers and single layer LSMO thin films at 10 K as a function of LSMO layer thickness.

Figure 7. Schematic model of the strain induced on each layer of the LSMO/PMN-PT bilayered heterostructures. The compressive and tensile stresses are indicated by the arrows.

Figure 8. Capacitance vs. frequency response under applied magnetic field of a single layer PMN-PT thin film at (a) 20K, (b) 100 K, (c) 200 K and (d) 300 K.

Figure 9. MC vs. magnetic field response of a single layer PMN-PT thin film at different temperatures.

Figure 10. Capacitance vs. frequency response under applied magnetic field of 140/20 bilayered heterostructure at (a) 20K, (b) 100 K, (c) 200 K and (d) 300 K.

Figure 11. MC (1T) vs. temperature response of LSMO/PMN-PT bilayered heterostructures.

Figure 12. Representative MC vs. magnetic field response of 140/20 bilayered heterostructure at 20 K. The solid line is a guide to the eye.